# Tailoring Topological Edge States with Photonic Crystal Nanobeam Cavities

Yongkang Gong[1], Liang Guo[1,2], Stephan Wong[1], Anthony J. Bennett[3], and Sang Soon Oh[1,*]

[1]*School of Physics and Astronomy, Cardiff University, Cardiff, CF24 3AA, United Kingdom.*
[2]*Department of Basic Science, Jilin Jianzhu University, 5088 Xincheng Street, Changchun, 130118, China.*
[3]*School of Engineering, Cardiff University, Cardiff, CF24 3AA, United Kingdom.*
*\*OhS2@cardiff.ac.uk*

**Abstract:** The realization of topological edge states (TESs) in photonic systems has provided unprecedented opportunities for manipulating light in novel manners. The Su-Schrieffer-Heeger (SSH) model has recently gained significant attention and has been exploited in a wide range of photonic platforms to create TESs. We develop a photonic topological insulator strategy based on SSH photonic crystal nanobeam cavities. In contrast to the conventional photonic SSH schemes which are based on alternately tuned coupling strength in one-dimensional lattice, our proposal provides higher flexibility and allows tailoring TESs by manipulating mode coupling in a two-dimensional manner. We reveal that the proposed hole-array based nanobeams in a dielectric membrane can selectively tailor single or double TESs in the telecommunication region by controlling the coupling strength of the adjacent SSH nanobeams in both vertical and horizontal directions. Our finding provides an in-depth understanding of the SSH model, and allows an additional degree of freedom in exploiting the SSH model for integrated topological photonic devices with unique properties and functionalities.

## 1. INTRODUCTION

In comparison with traditional photonic defect states that are sensitive to perturbations, edge states from photonic topological insulators (PTIs) are robust against local perturbations and immune to back scattering. This leads to intriguing and unexpected photonic devices and functionalities for electromagnetic wave manipulations, such as unidirectional light and backscattering-free light transport [1-3], topological lasing [4-11], light steering [12], nonlinear parametric generation [13, 14], protection of single photons [15], and entangled photonic states [16, 17]. The generation of topological edge states (TESs), which are the core of the emerging field of photonic PTIs, has recently made remarkable progress and has inspired a number of fundamentally different topological approaches. For example, a photonic analogue of a quantum Hall topological insulator was developed in the microwave regime using gyromagnetic materials with a strong magnetic field applied to break the time-reversal symmetry, and unidirectional backscattering-immune TESs were observed [18]. Later, a number of proposals have been put forward to realize TESs free of external magnetic fields by temporal modulation of photonic crystals to mimic time-reversal-symmetry breaking [19-22]. All-dielectric PTI approaches based on pseudo-time-reversal symmetry [15, 23-28] and valley Hall photonic crystals with broken spatial-inversion symmetry [2, 3, 29-37] have proven to be effective in generating TESs at the subwavelength scale. Another elegant and powerful subwavelength-scale nontrivial topology approach is the Su-Schrieffer-Heeger (SSH) model, which was originally introduced to describe fractionalized electric charges in polyacetylene and has recently attracted considerable attention in photonics. Photonic SSH structures have been extensively investigated and applied to a broad range of platforms from microwave to optical regime including plasmonic waveguides [38], zigzag arrays of dielectric resonator chain [39], dielectric nanoparticles [40] , polariton micropillars [7], micro-ring resonator arrays [41, 42], photonic crystal L3 nanocavity dimer array [43], dielectric waveguides [44-48]. These platforms open avenues to on-chip photonic devices for robust topologically protected light manipulation.

    In this paper, we propose a new SSH scheme based on photonic crystal (PhC) nanobeams and demonstrate that TESs can be generated by controlling the coupling strength of the nanobeams in two dimensions, which differs from the reported photonic SSH structures that utilize alternate modulation of the coupling strength in a one-dimensional lattice. The SSH nanobeams can allow two types of TESs in the telecommunication wavelength region, and more importantly the TESs can be selectively enabled by engineering the vertical spacing and horizontal shift between the adjacent nanobeams in the SSH structures. The proposed SSH nanobeam concept provides a new kind of high Q factor and integrated photonic topological platform, which we believe could find promising applications for various passive and active topological devices and functionalities such as topological lasing and single photon emission.

## 2. CONCEPT, IMPLEMENTATION AND ANALYSIS OF THE SSH NANOBEAMS

Our proposed SSH scheme utilizes a free-standing PhC nanobeam array with each array consisting of a row of air holes in a semiconductor membrane with thickness $t$ and width $w$. Since the topological property of the proposed

SSH nanobeam structure arises from the alternating coupling strength between the adjacent nanobeams, we first investigate the optical coupling characteristic of two identical nanobeams, as schematically depicted in Fig. 1(a), where the two nanobeams have vertical spacing of $d_1$ and horizontal shift of $d_2$ with each nanobeam incorporating six air holes in the reflector sections and nine air holes in the taper section. All the holes have the same diameter. The hole-to-hole spacing in the two reflector sections is the same but reduces gradually from both sides to the center of the taper section to form an optical cavity. We perform designs and analyses by three-dimensional (3D) finite-difference time-domain (FDTD) method [49]. The obtained results demonstrate that when the two nanobeams have a large vertical spacing such as $d_1 = 2$ μm, each nanobeam generates two resonance modes in the telecommunication region and the resonance modes in the two nanobeams do not couple to each other, as illustrated by the electric field distribution $|E|$ in Fig. 1(c). We note from the optical spectrum in Fig. 1(b) and the field distribution $H_z$ in Fig. A1 in Appendix A that the first resonance mode at $\lambda_s = 1.546$ μm has symmetric $H_z$ field distribution with respect to the $x = 0$ μm plane, while the second resonance mode at $\lambda_a = 1.624$ μm has antisymmetric $H_z$ field profile with respect to the same plane. The symmetric and antisymmetric modes have Q factor of $7.1 \times 10^4$ and $1.1 \times 10^4$, and mode volume $V_{\text{eff}}$ of $0.5(\lambda/n)^3$ and $0.9(\lambda/n)^3$, respectively, where $\lambda$ is the resonance mode wavelength and $n$ is the refractive index of the semiconductor membrane. The mode volume is calculated by $V_{\text{eff}} = \frac{\int_V \varepsilon(r)|E(r)|^2 d^3(r)}{\max\left[\varepsilon(r)|E(r)|^2\right]}$ [50], where $\varepsilon(r)$, $|E(r)|$, and $V$ are the dielectric constant, the electric field strength, and the volume of the nanobeam cavity, respectively. We stress that the Q factor and the mode volume can be further improved by optimizing the hole diameters, the number of holes, and the hole-to-hole spacings as reported in Refs. [51-57]. Whilst the two resonance modes in each nanobeam do not couple to each other because they are separated at large vertical spacing of $d_1 = 2$ μm (Fig. 1(c)), they start to couple when $d_1$ reduces to such as $d_1 = 0.6$ μm (Fig. 1(d)). As a result, wavelength splitting of the two resonance modes occurs. We see from the optical spectra (Fig. 1(b)) and the field distribution $H_z$ (Fig. A2 in Appendix A2) that the first resonance mode at $\lambda_s$, which is symmetric with respect to reflection by the $x = 0$ plane, is split into antisymmetric and symmetric modes with respect to reflection by the $y = 0$ μm plane. In the optical spectrum, the antisymmetric (symmetric) mode is located at shorter (longer) wavelength of $\lambda_{s,a}$ ($\lambda_{s,s}$). A similar phenomenon happens to the second resonance mode at $\lambda_a$ which is split into an antisymmetric mode with respect to the reflection by the $y = 0$ μm plane at shorter wavelength of $\lambda_{a,a}$ and a symmetric mode with respect to the reflection by the $y = 0$ μm at longer wavelength of $\lambda_{a,s}$. The wavelength splitting strongly depends on the vertical spacing $d_1$ (Fig. 1(e)) and increases when $d_1$ decreases due to the presence of stronger mode coupling. We derive the coupling strength $k$ of the two resonance modes by $\kappa = \frac{\pi c \Delta \lambda}{\lambda^2}$ [58], where $c$ is the speed of light in vacuum, and $\Delta \lambda = |\lambda_{s,a} - \lambda_{s,s}|$ ( $\Delta \lambda = |\lambda_{a,a} - \lambda_{a,s}|$) and $\lambda = \lambda_s$ ($\lambda = \lambda_a$) is the splitting wavelength difference and the wavelength of the first (second) resonance mode, respectively. Figure 1(f) shows that the first resonance mode has lower coupling strength than that of the second resonance mode under the same $d_1$ due to smaller mode volume.

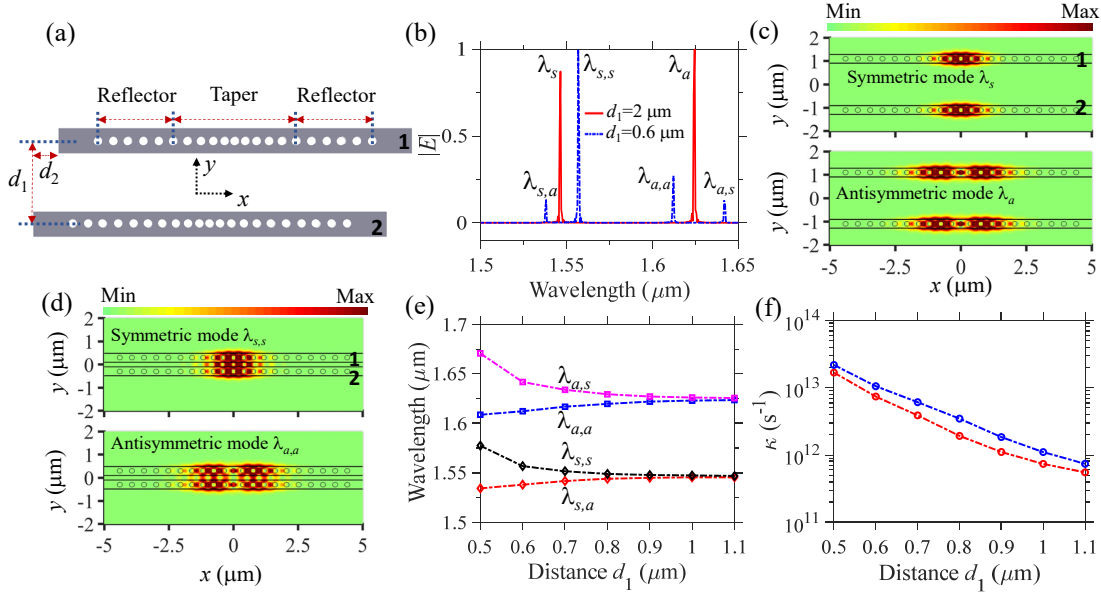

**Fig. 1**. Wavelength splitting characteristics of the resonance modes of the two coupled PhC nanobeams. (a) Schematic top view of the two identical nanobeams with vertical spacing $d_1$ and horizontal shift $d_2$, where each nanobeam consists of a linear array of air holes with the same radius of $r$ in a semiconductor membrane with thickness $t$ and width $w$. (b) The normalized $|E|$ spectra of the nanobeams when $d_1 = 2$ μm and $d_1 = 0.6$ μm, showing the appearance of the wavelength splitting phenomenon at small $d_1$ due to strong mode coupling. (c) The electric field distribution $|E|$ of the symmetric mode at $\lambda_s = 1.546$ μm and the antisymmetric mode at $\lambda_a = 1.624$ μm in the plane of the central membrane (i.e., the $z = 0$ μm plane) when $d_1 = 2$ μm, respectively. (d) The field distribution $|E|$ of the symmetric mode at $\lambda_{s,s} = 1.557$ μm and the antisymmetric mode at $\lambda_{a,a} = 1.612$ μm when $d_1 = 0.6$ μm, respectively. The field distribution $H_z$ of these modes are given in Figs. 1A and 2A in Appendix A. (e) Evolution of the four splitting wavelengths (i.e., $\lambda_{s,a}$, $\lambda_{s,s}$, $\lambda_{a,a}$, and $\lambda_{a,s}$) to the change of $d_1$. (f) Dependence of the coupling strength $\kappa$ of the first (red curve) and the second resonance modes (blue curve) on $d_1$. In above simulations, each nanobeam has six air holes with hole-to-hole spacing of 0.47 μm in the reflector sections, and has nine nanoholes in the taper section with the hole-to-hole spacing decreased with a step of 20 nm from either side of the nanobeam to its center. The other geometrical parameters are $t = 0.3$ μm, $w = 0.4$ μm, $r = 0.1$ μm, and $d_2 = 0$ μm. The refractive index of the semiconductor membrane is considered to be 3.3.

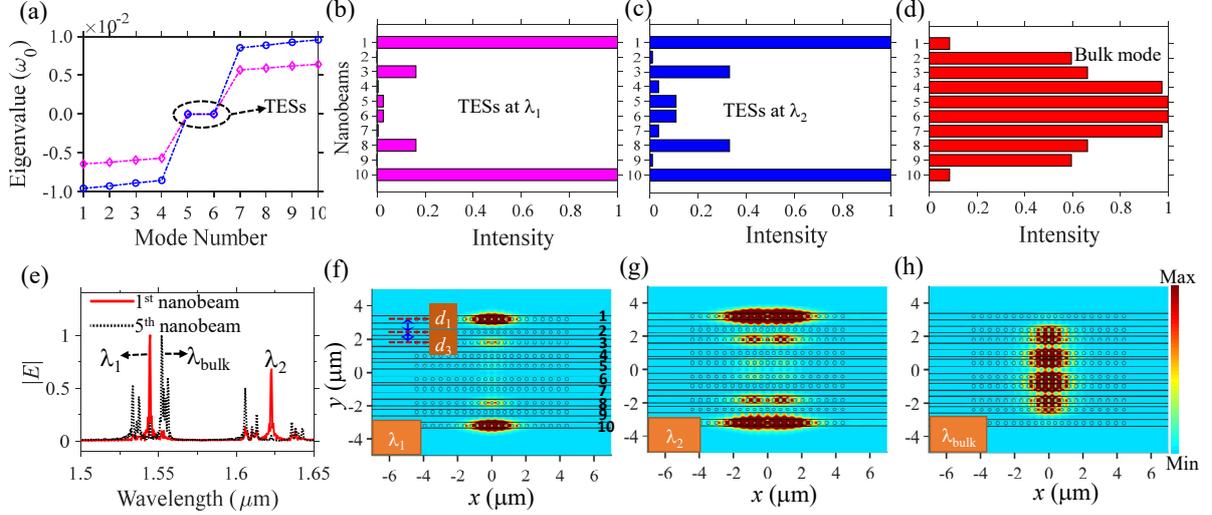

**Fig. 2**. Generation of TESs with the finite SSH nanobeams implemented by the tight-binding model ((a)-(d)) and the 3D FDTD method ((e)-(h)). (a) The eigenvalues normalized to the angular frequency for the first (the magenta curve) and the second (the blue curve) resonance modes, indicating the existence of the two degenerated zero-energy TESs at both the first resonance wavelength $\lambda_1$ and the second resonance wavelength $\lambda_2$. The field distribution (i.e., the real part of the eigenvectors) of these TESs have localized field at edge nanobeams in symmetric or antisymmetric manners, as plotted in Fig. A4 in Appendix B. (b)-(d) The normalized intensity (i.e., the absolute value of the eigenvector) of the TESs at $\lambda_1$, $\lambda_2$, and one of the bulk modes, respectively. (e) The normalized $|E|$ spectra of the edge nanobeam (i.e., the 1st nanobeam) and the middle nanobeam (i.e., the 5th nanobeam) of the SSH structure. The first (fifth) nanobeam has the same spectra to that of the tenth (sixth) nanobeam due to structure symmetry. (f)-(h) The electric field distribution $|E|$ of the first TES at $\lambda_1 = 1.546$ µm, the second TES at $\lambda_2 = 1.624$ µm, and the bulk mode at $\lambda_{bulk} = 1.552$ µm in the middle plane of the nanobeam membrane, respectively. The SSH structures under study have 10 identical nanobeams with alternative neighboring distance of $d_1 = 0.8$ µm and $d_3 = 0.6$ µm, and zero horizontal shift (i.e., $d_2 = 0$ µm), as denoted by the structure outline in (f). Each nanobeam of the SSH structure has geometrical parameters same as those in Fig. 1.

Based on the property of the optical coupling strength varying with the nanobeams' vertical spacing, we construct finite SSH nanobeams to generate TESs by alternately changing the vertical spacing between successive nanobeams. For the examples considered here we use structures with 10 nanobeams, but the results have general applicability to other even-number nanobeam arrays. When our SSH structures have an odd number of nanobeams, the generated TESs have the similar physical mechanism and optical properties (although there is a difference that odd number of SSH nanobeams allows a single zero-energy mode while an even number of nanobeams generates two degenerate zero-energy modes [59]). The developed SSH structure involves a nontrivial termination with spacing of $d_1 = 0.8$ µm and $d_3 = 0.6$ µm (as denoted by the structure outline in Fig. 2(f)), which gives rise to alternative intra-coupling of $\kappa_1 = 5.5 \times 10^{11}$ $s^{-1}$ and inter-coupling of $\kappa_2 = 7.3 \times 10^{12}$ $s^{-1}$ for the first resonance mode, and alternative intra-coupling of $\kappa_1 = 7.4 \times 10^{11}$ $s^{-1}$ and inter-coupling of $\kappa_2 = 1.1 \times 10^{13}$ $s^{-1}$ for the second resonance mode. We perform tight binding analysis by incorporating the above coupling strength into the Hamiltonian of the SSH system (see Eq. A1 in Appendix B) to obtain the eigenvalues and the eigenvectors. Figures 2(a) demonstrates that the SSH configuration allows two degenerate zero-energy modes at both the first resonance wavelength $\lambda_1$ and the second resonance wavelength $\lambda_2$. At each of the resonance wavelengths, one zero-energy mode has symmetric localized field at the edge nanobeams (i.e., the first and tenth nanobeams), while the other zero-energy mode has an asymmetric localized field profile (Fig. A4 in the Appendix B). The intensity of the TESs at $\lambda_1$ and $\lambda_2$ are mainly localized at the center of the edge nanobeams and decay exponentially to the middle nanobeams, as depicted in Figs. 2(b) and 2(c). Conversely, bulk modes do not have localized fields in the edge nanobeams and have fields mainly distributed in the middle nanobeams (Fig. 2(d)). We validate the tight binding analysis by implementing 3D FDTD modellings. Figure 2(e) demonstrates that the first nanobeam supports two strong spectral peaks at wavelengths of $\lambda_1 = 1.546$ µm and $\lambda_2 = 1.642$ µm, which corresponds to the two TESs of the SSH structures with electric field mainly localized at the first and tenth nanobeams (Figs. 2(f) and 2(g)). We note that from Fig. 2(e) that the fifth nanobeam supports multiple spectral peaks with wavelengths different to that of the TESs. These peaks correspond to bulk modes which allow strong electric field distributed at the second to the ninth nanobeams (Fig. 2(h)). It is clearly that the FDTD results are in good agreement with our analytical tight binding analysis. The field strength of the TESs is determined by an exponential decay function $\left(\frac{\kappa_2}{\kappa_1}\right)^{-n}$ [41, 43], where $\kappa_1$ ($\kappa_2$) represents the intra (inter) coupling strength, $n$ ($n = 1, 2, ... N/2$) stands for the $n$-th nanobeam in the SSH structure, and $N$ denotes the total number of the nanobeams and is even integer. For example, when we increase the intra-coupling strength $\kappa_1$ by reducing $d_1$ from 0.8 μm to 0.7 μm while keeping the inter-coupling strength $\kappa_2$ by keeping $d_3 = 0.6$ μm, as shown in Fig. A5 in Appendix B, the exponential decay function increases and thus the electric field decays slower from the edge to the middle nanobeams. It is noted from Figs. 2(b) and 2(c) that the first TES decays faster from edge nanobeams to middle nanobeams that that of the second TES, since it has a larger exponential decay function. In our FDTD simulations, multiple electric dipoles are placed randomly in each nanobeam to excite all the possible resonance modes. Meanwhile, several monitors are added randomly in each nanobeam to record the time signal to retrieve the averaged optical spectra.

## 3. TAILORING TESs VIA TUNING COUPLING STRENGTH BOTH VERTICALLY AND HORIZONTALLY

Having established the underlying concept of the topological SSH nanobeams by alternatively varying the vertical spacing between the adjacent nanobeams, we continue to explore another interesting aspect of selectively manipulating the TESs by controlling the horizontal shift of the successive nanobeams. We investigate the wavelength splitting of the resonance modes of the two coupled nanobeams when their horizontal shift $d_2$ changes at a fixed vertical spacing $d_1$ (see Fig. 1(a)). Figure 3(a) demonstrates that the four splitting wavelengths vary quasi-periodically with $d_2$, which differ significantly from the wavelength splitting behavior in Fig. 1(e). We observe that when $d_2$ changes from 0 μm to 0.2 μm, the splitting wavelengths of $\lambda_{11}$ ($\lambda_{21}$) move closer to its counterparts of $\lambda_{12}$ ($\lambda_{22}$), meaning the coupling strength of the two resonance modes supported by the two nanobeams becomes weaker. We can get insight into this optical property by looking at the field distributions at these splitting wavelengths. The air holes in the center of the two nanobeams have no horizontal shift to each other when $d_2 = 0$ μm (Fig. 3(b)), and maximum optical coupling and maximal wavelength splitting occur (Fig. 3(a)). When $d_2$ increases to 0.2 μm, however, the air holes in the center of the two nanobeams are shifted by nearly half pitch to each other (Fig. 3(c)), resulting in small optical coupling and smaller wavelength splitting. When $d_2$ is further enlarged to 0.4 μm (Fig. 3(d)), the air holes in the cavity center of the two nanobeams are shifted by nearly one pitch to each other, and the mode coupling and wavelength splitting becomes stronger again.

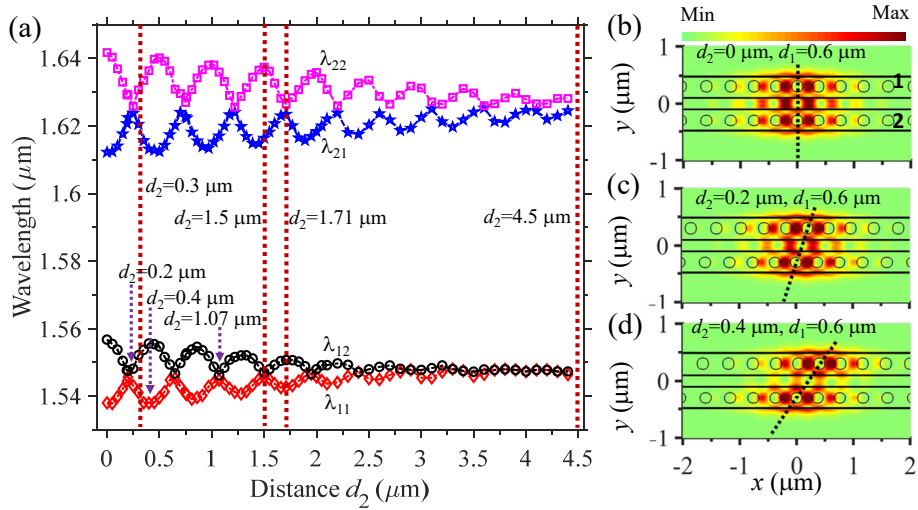

**Fig. 3**. Influence of the horizontal shift $d_2$ to the wavelength splitting of the resonance modes of the double coupled-nanobeams (see Fig. 1(a)). (a) The dependence of the four splitting wavelengths (i.e., $\lambda_{11}$, $\lambda_{12}$, $\lambda_{21}$ and $\lambda_{22}$) on $d_2$, which indicates quasi-periodic change of the coupling strength of the resonance modes and hence allows us to selectively generate TESs by the finite SSH nanobeams. (b)-(d) Zoom of the electric field distribution $|E|$ in the middle plane of the nanobeam membrane (also see field distribution $H_z$ in Fig. A3 in Appendix A) at the splitting wavelengths of $\lambda_{11} = 1.538$ μm when $d_2 = 0$ μm, $\lambda_{11} = 1.546$ μm when $d_2 = 0.2$ μm, and $\lambda_{11} = 1.538$ μm when $d_2 = 0.4$ μm, respectively, where the black dotted lines link the central holes of the two nanobeams. The vertical spacing in above simulations is fixed at $d_1 = 0.6$ μm and the other geometrical parameters are the same as those in Fig. 1.

Another intriguing feature is that the wavelength splitting of the first resonance mode (i.e., $\lambda_{11}$ and $\lambda_{12}$) and the second resonance mode (i.e., $\lambda_{21}$ and $\lambda_{22}$) respond differently to the change of $d_2$ due to their different field distributions. Since the first resonance mode has smaller mode volume and hence smaller field spreading in the horizontal direction, its mode coupling tends to change faster when $d_2$ varies. It is found that the wavelength splitting of the first resonance mode is smaller than that of the second resonance mode when $d_2 = 1.5$ μm, but becomes larger than that of the second resonance mode when $d_2 = 1.71$ μm as depicted by the vertical dotted lines in Fig. 3(a). This unique feature allows us to selectively enable TESs by the finite SSH nanobeams. When the horizontal shift is as large as $d_2 = 4.5$ μm, the resonance modes in the two nanobeams are separated far away from each other and consequently mode coupling no longer occurs.

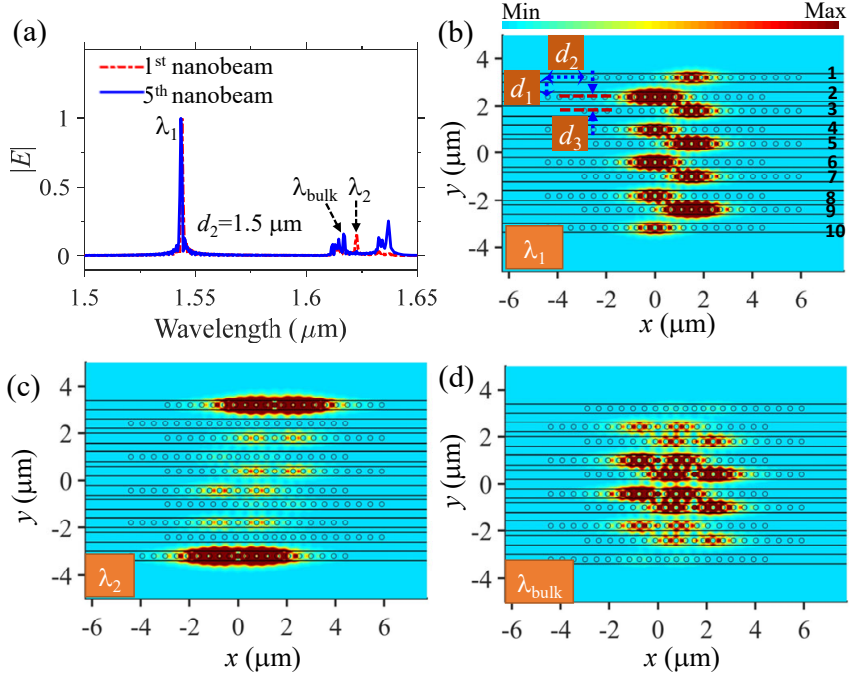

**Fig. 4**. Optical properties of the TESs enabled by SSH configuration with horizontal shift of $d_2 = 1.5$ μm between the adjacent nanobeams. (a) The spectra of the first and the fifth nanobeams of the SSH structure, showing a merged peak at wavelength $\lambda_1 = 1.546$ μm and suggesting the edge mode not being allowed at $\lambda_1$ due to weak coupling strength at this wavelength. (b)-(d) The electric field distribution $|E|$ of the SSH structure in the middle plane of the nanobeam membrane at $\lambda_1 = 1.546$ μm, $\lambda_2 = 1.624$ μm, $\lambda_{bulk} = 1.617$ μm, respectively. The geometrical parameters of the SHH structures are the same as those in Fig. 2 except a 1.5 μm horizontal shift between the neighboring nanobeams.

Given the fact that the coupling strength of the resonance modes depends on the horizontal shift of the adjacent nanobeams as just revealed, another degree of freedom can be added to tailor TESs. It is possible to selectively generate either of the TESs by carefully shifting nanobeams horizontally. We observe from Fig. 4 that the resonant mode at $\lambda_1 = 1.546$ μm is no longer an edge mode when the horizontal shift changes from $d_2 = 0$ μm to $d_2 = 1.5$ μm. In this scenario, the first resonance mode has low coupling strength (Fig. 3(a)) and does not support strong alternative intra and inter coupling between the adjacent nanobeams even though the structure has SSH configuration. As a result, the resonant mode at $\lambda_1$ is excited in each nanobeam cavity, and no edge states appear for $\lambda_1$ as indicated by the optical spectrum (Fig. 4(a)), where both the edge and middle nanobeams share the same spectra peak wavelength at $\lambda_1$. This is evidenced by the field profile plotted in Fig. 4(b), where the electric field is distributed along all of the ten nanobeams. When $d_2 = 1.5$ μm, TES appears only for the second resonance mode at $\lambda_2 = 1.624$ μm because of strong light coupling between the neighboring nanobeams only at this wavelength. The edge nanobeams (i.e., the first and tenth nanobeams) have a spectral peak at $\lambda_2$, and support strong localized electric field along them (Fig. 4(c)). The field distribution of the bulk mode at $\lambda_{\text{bulk}} = 1.615$ μm is plotted in Fig. 4(d). We see that the bulk mode has electric field mainly distributed at second to ninth nanobeams, which differs to the field profiles of both the resonant mode at $\lambda_1$ and the edge mode at $\lambda_2$. Based on the same concept, we can select the first resonance mode to become TES and at the same time disable the topological edge state property of the second resonance mode by changing the horizontal shift to $d_2 = 1.71$ μm (Figs. 5(a) and (b)), since in this case the first resonance mode supports high coupling strength while the second resonance mode has small coupling strength as suggested by Fig. 3(a). When the horizontal shift is as large as $d_2 = 4.5$ μm in the SSH configuration, both resonance modes have weak coupling strength and thus do not support edge modes any more, as demonstrated by the spectra in Fig. 5(c) and the field distribution in Fig. 5(d).

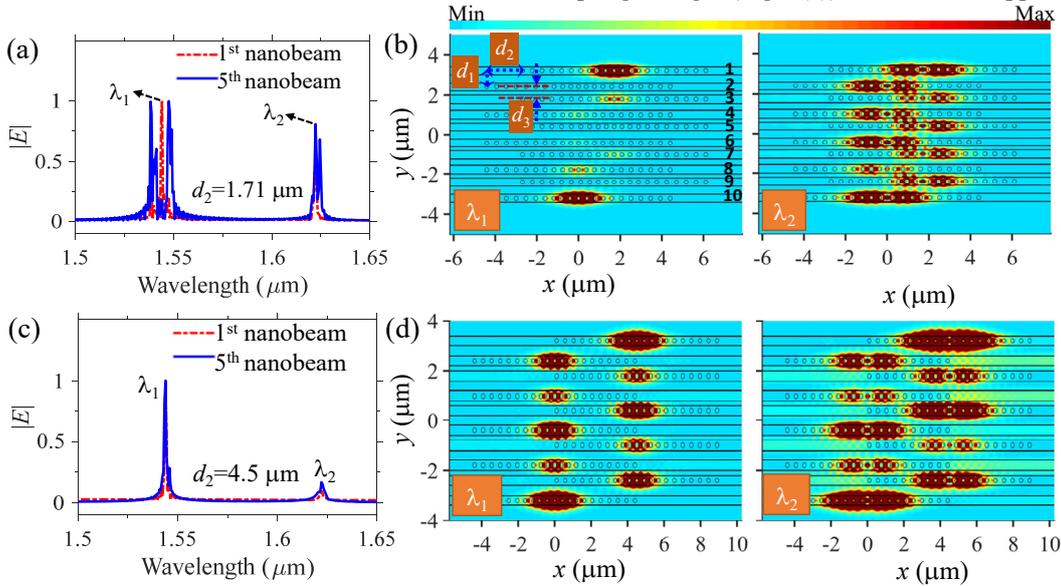

**Fig. 5**. Tailoring TESs with different horizontal shift of the SSH nanobeams. (a) Spectra of the edge and middle nanobeams of the SSH structure when $d_2 = 1.71$ μm, showing a merged peak at $\lambda_2 = 1.624$ μm and suggesting the edge mode not being allowed at this wavelength. (b) The corresponding electric field distribution $|E|$ at the spectra peak wavelengths $\lambda_1$ (left) and $\lambda_2$ (right), respectively. (c) Spectra of the edge and middle nanobeams of the SSH structure when $d_2 = 4.5$ μm, showing merged peaks at both $\lambda_1 = 1.546$ μm and $\lambda_2 = 1.624$ μm and indicating non generation any edge mode. (d) The corresponding electric field distribution $|E|$ at the two wavelengths. The other geometrical parameters of the nanobeams are the same as those in Fig. 4.

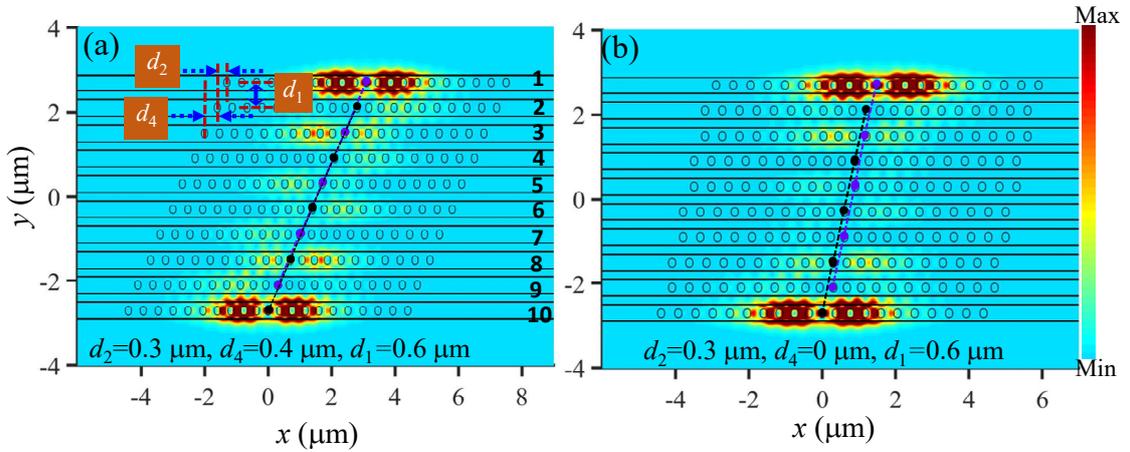

**Fig. 6**. Generation of TESs by the SSH nanobeams with the same vertical spacing but different horizontal shift. (a) The field distribution $|E|$ of the edge mode at $\lambda_2 = 1.624$ μm when $d_2 = 0.3$ μm and $d_4 = 0.4$ μm. (b) The field distribution $|E|$ of the edge mode at $\lambda_2 = 1.624$ μm when $d_2 = 0.3$ μm and $d_4 = 0$ μm. In the SSH structures, all the nanobeams have the same vertical spacing of $d_1 = 0.6$ μm. The horizontal shift between the nanobeams $j$ to $j + 1$ ($j = 1, 3, 5, 7, 9$) is $d_2$, and the horizontal shift between the nanobeams $j$ to $j +1$ ($j =2, 4, 6, 8$) is $d_4$, as depicted by the structure outline in (a). The other geometrical parameters of the nanobeams are the same as those in Fig. 5. The black and purple dotted lines indicate the central holes of the even and odd nanobeams, respectively. The spectra of the SSH nanobeams and the field distribution $|E|$ of the generated edge mode at the first resonance mode wavelength $\lambda_1$ are given in Fig. A6 in Appendix C.

Lastly, we exploit another kind of SSH nanobeam structure to enable TESs using the same vertical spacing but different horizontal shift between the adjacent nanobeams. For example, two edge modes can be generated when the horizontal shift between the nanobeams $j$ to $j + 1$ ($j = 1, 3, 5, 7, 9$) is $d_2 = 0.3$ μm, the horizontal shift between the nanobeams $j$ to $j + 1$ ($j = 2, 4, 6, 8, 10$) is $d_4 = 0.4$ μm, and all the nanobeams have the same vertical spacing of $d_1 = 0.6$ μm, as shown in Fig. 6(a). This topological property is attributed to the mode coupling strength between the neighboring nanobeams being smaller at $d_2 = 0.3$ μm than that at $d_4 = 0.4$ μm under the same the horizontal shift $d_1$ (Fig. 3(a)), which guarantees the alternative intra and inter coupling required by the SSH model. Based on the same working principle, we can also enable TESs when $d_2 = 0.3$ μm an $d_4 = 0$ μm. Figures 6(a) and 6(b) plot the electric field distribution of the generated edge mode at $\lambda_2 = 1.624$ μm.

## 4.  CONCLUSION

To conclude, we have demonstrated a PTI strategy based on the finite SSH nanobeams with PhC nanohole arrays in a semiconductor membrane. The key advantage of the developed SSH architecture is that it allows manipulation of TESs in two-dimensional manner, which differs to the reported SSH structures that generally use one-dimensional photonic lattice. We have designed SSH nanobeams with various configurations, and have investigated the coupling strength, the spectral characteristics and the field distributions of the TESs. The proposed structures can allow two edge states with different mode profiles in the telecommunication region. Moreover, we reveal that the two edge states can be flexibly controlled by tailoring the mode coupling strength. It is noted that a single edge state can be selectively chosen by carefully shifting nanobeams vertically and horizontally. We believe our findings deepen the understanding of the SSH model for TESs generation and manipulation, and provide a new integrated photonic strategy to retrieve nontrivial topology.

## APPENDIX A: WAVELENGTH SPLITTING OF TWO COUPLED NANOBEAM CAVITIES

The proposed single nanobeam cavity supports two high Q factor resonance modes in the telecommunication region. The first resonance mode has wavelength of $\lambda_s = 1.546$ μm with symmetrical $H_z$ field distribution with respect to the $x = 0$ μm plane (Fig. A1(a)), and the second resonance mode is at wavelength of $\lambda_a = 1.624$ μm and has antisymmetric $H_z$ field profile along the $x = 0$ μm plane (Fig. A1(b)). When the vertical spacing of the two coupled nanobeams is as large as $d_1 = 2$ μm, the two resonance modes have localized field at the center of each nanobeam and the modes from the two nanobeams do not couple to each other. When the spacing $d_1$ reduces to 0.6 μm, strong mode coupling occurs and the two resonance modes are split (Fig. 1(e)). The first

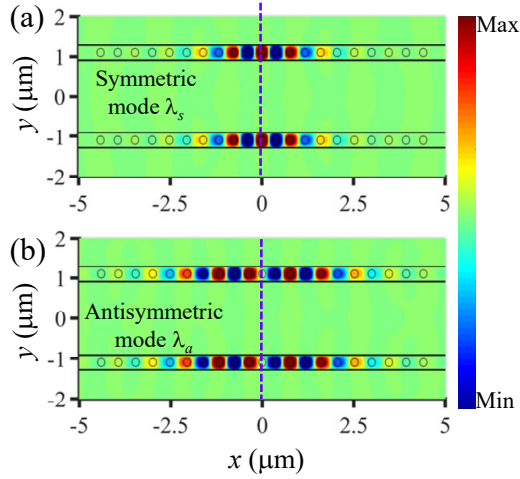

**Fig. A1.** The real part of the field distribution $H_z$ of the symmetric mode at $\lambda_s = 1.546$ μm (a) and the antisymmetric mode at $\lambda_a = 1.624$ μm in the plane of the central membrane (i.e., the $z = 0$ μm plane). The geometrical parameters of the nanobeams are the same as those in Fig. 1(c). The purple dashed lines indicate the $x = 0$ μm plane.

resonance mode is split into a mode at wavelength of $\lambda_{s,a}$ and a mode at wavelength of $\lambda_{s,s}$. The mode at $\lambda_{s,a}$ has symmetric (antisymmetric) $H_z$ field profile with respect to the $x = 0$ μm ($y = 0$ μm) plane, while the mode at $\lambda_{s,s}$ has symmetric $H_z$ field profile with respect to both the $x = 0$ μm and the $y = 0$ μm planes, as depicted in Figs. A2(a) and A2 (b). The second resonance mode is split into two modes at $\lambda_{a,a}$ and $\lambda_{a,s}$. Figures A2(c) and A2(d) indicates that the mode at $\lambda_{a,a}$ has antisymmetric $H_z$ field profile with respect to both the $x = 0$ μm and the

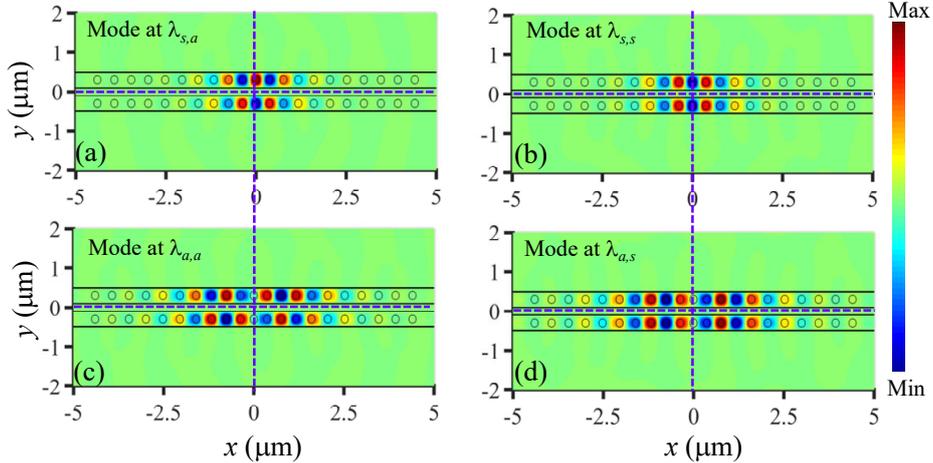

**Fig. A2** The real part of the field distribution $H_z$ of the splitting modes at $\lambda_{s,a} = 1.538$ μm (a), $\lambda_{s,s} = 1.557$ μm (b), $\lambda_{a,a} = 1.612$ μm (c), and $\lambda_{a,s} = 1.642$ μm (d), respectively. The geometrical parameters of the nanobeams are the same as those in Fig. 1(d). The purple dashed lines indicate the $x = 0$ μm and the $y = 0$ μm planes.

$y = 0$ μm planes, while the mode at $\lambda_{a,s}$ has antisymmetric (symmetric) $H_z$ field profile with respect to the $x = 0$ μm ($y = 0$ μm) plane. The horizontal spacing $d_2$ between the two nanobeams affect the coupling strength of the resonance modes as well. The mode coupling strength changes quasi-periodically with $d_2$ (Fig. 3(a)). Figures 3(c) and 3(d) show the field distribution $|E|$ at the splitting wavelength of 1.546 μm when $d_2 = 0.2$ μm and at the splitting wavelength of 1.538 μm when $d_2 = 0.4$ μm. Figure. A3 depicts their corresponding $H_z$ field distribution.

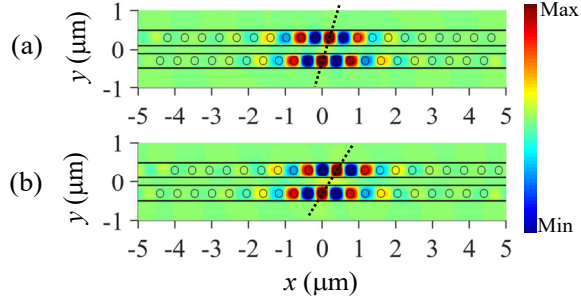

**Fig. A3**. The real part of the field distribution $H_z$ of the two coupled nanobeams at the splitting wavelength of 1.546 μm when $d_2 = 0.2$ μm (a) and at the splitting wavelength of 1.538 μm when $d_2 = 0.4$ μm (b), respectively. The geometrical parameters of the nanobeams are the same as those in Figs. 3(c) and 3(d). The black dotted lines link the center of the two nanobeams.

## APPENDIX B: HAMILTONIAN AND TIGHT BINDING ANALYSIS

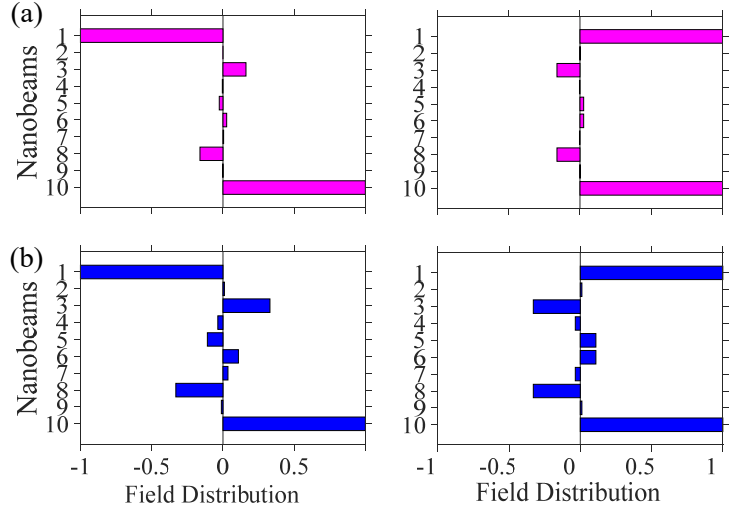

**Fig. A4**. The normalized field distribution (the real part of the eigenvectors) of the two degenerate zero-energy TESs at the first resonance wavelength $\lambda_1$ (a) and the second resonance wavelength $\lambda_2$ (b) obtained by the tight binding model. The simulation paraments are the same as those in Figs. 2(a)-(c).

The proposed SSH structure has ten nanobeams with alternatively changed spacing between the successive nanobeams, as indicated by structure outline in Fig. 2(f). With the nearest-neighbor approximation, the Hamiltonian of the finite SSH nanobeams can be given by [41]

$$H = \begin{pmatrix} 0 & \kappa_1 & & & & \\ \kappa_1 & 0 & \kappa_2 & & & \\ & \kappa_2 & \ddots & \kappa_1 & & \\ & & \kappa_1 & 0 & \kappa_2 & \\ & & & \kappa_2 & 0 & \end{pmatrix}_{N \times N}, \qquad \text{(A1)}$$

where $N$ is the total number of the nanobeams, and $\kappa_1$ ($\kappa_2$) is the intracell (intercell) coupling strength. The diagonal terms can be viewed as the detuning with respect to resonant frequency. By including the obtained coupling strength (see Fig. 1(f)) in above equation, we can derive the eigenvalues and the eigenvectors of the proposed SSH nanobeam structures. The field distribution of the two zero-energy TESs at the first and second resonance mode wavelengths have localized field at edge nanobeams in symmetric or antisymmetric manners, as shown in Fig. A4. Figure 2 and Fig. A5 demonstrate that our tight binding analysis agrees well with the FDTD results.

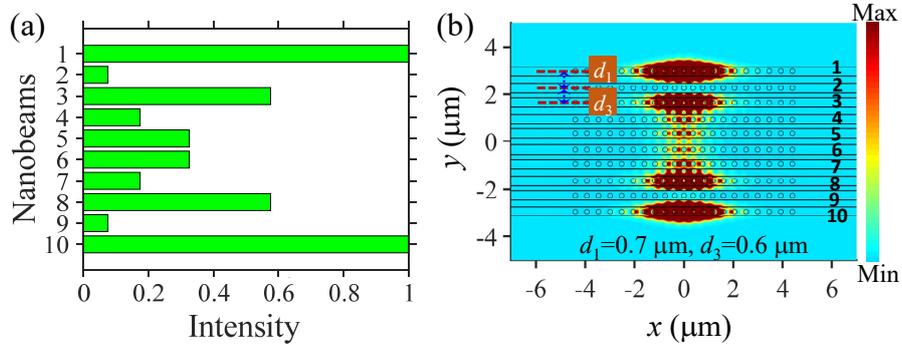

**Fig. A5**. The normalized intensity (a) and the field distribution $|E|$ (b) of the first TES obtained by the tight bonding model and the FDTD method. Here, $d_1 = 0.7$ μm, $d_3 = 0.6$ μm, and the other geometrical parameters are the same as those in Fig. 2.

## APPENDIX C: OPTICAL PROPERTIES OF THE SSH NANOBEAMS WITH THE SAME VERTICAL SPACING

Since mode coupling strengthen can be tuned by the horizontal shift of the nanobeams (Fig. 3(a)), we can tailor TESs with SSH structures having the same vertical spacing but different horizontal shift between the adjacent nanobeams, as demonstrated in Fig. 6 and A6.

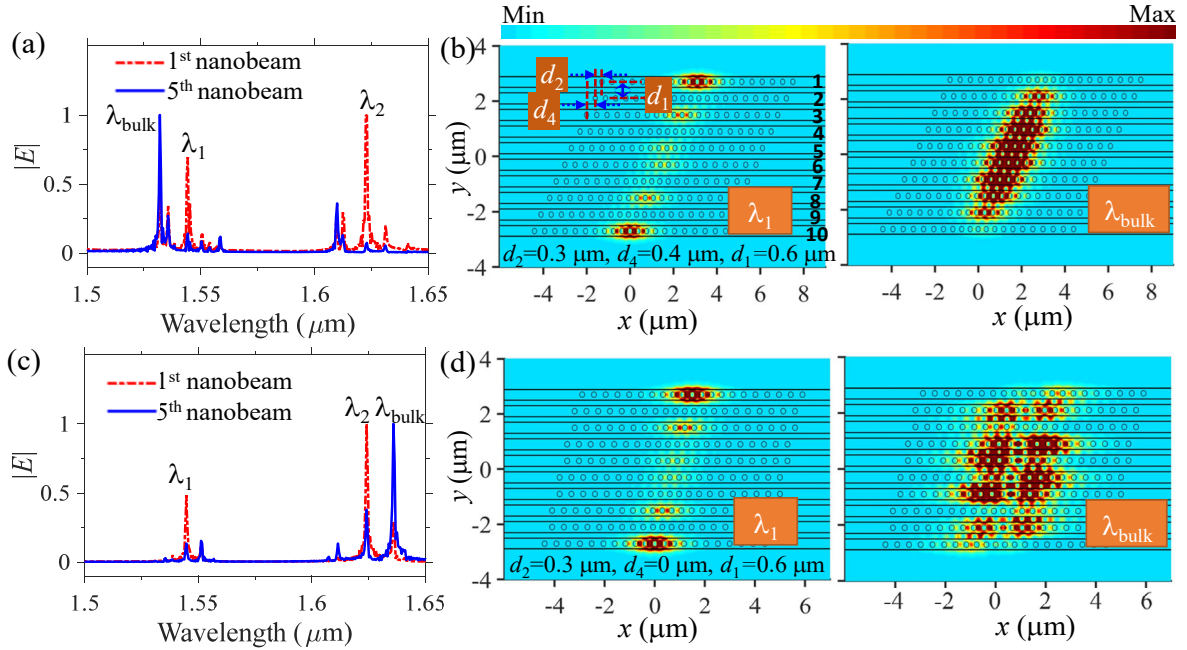

**Fig. A6.** (a) The optical spectra of the edge and middle nanobeams when $d_2 = 0.3$ μm and $d_4 = 0.4$ μm. (b) The corresponding field distribution $|E|$ of the edge mode at $\lambda_1 = 1.546$ μm (left figure) and the bulk mode at $\lambda_{\text{bulk}} = 1.532$ μm (right figure). (c) The optical spectra of the edge and middle nanobeams when $d_2 = 0.3$ μm and $d_4 = 0$ μm. (d) The corresponding field distribution $|E|$ of the edge mode at $\lambda_1 = 1.546$ μm (left figure) and the bulk mode at $\lambda_{\text{bulk}} = 1.636$ μm (right figure). All the adjacent nanobeams have the same vertical spacing of $d_1 = 0.6$ μm. The geometrical paraments of SSH structures are the same as those in Fig. 6.

**Funding.** The work is part-funded by the European Regional Development Fund through the Welsh Government (80762-CU145 (East)).

**Disclosures**
The authors declare no conflicts of interest.